\documentclass[twocolumn,trackchanges]{aastex62} 

\submitjournal{ApJL}

\shorttitle{Evolution of dark energy reconstructed from the latest observations}
\shortauthors{Wang et al.}

\def\ie{{\frenchspacing\it i.e.}}
\def\eg{{\frenchspacing\it e.g.~}}
\def\etc{{\frenchspacing\it etc.}}

\def\be{\begin{equation}}
\def\ee{\end{equation}}
\def\ba{\begin{eqnarray}}
\def\ea{\end{eqnarray}}

\begin{document}

\title{Evolution of dark energy reconstructed from the latest observations}

\correspondingauthor{Yuting Wang}
\email{ytwang@nao.cas.cn}

\author[0000-0001-7756-8479]{Yuting Wang}
\affiliation{National Astronomy Observatories, Chinese Academy of Science, Beijing, 100101, P.R.China}

\author{Levon Pogosian}
\affiliation{Department of Physics, Simon Fraser University, Burnaby, British Columbia, Canada V5A 1S6}
\affiliation{Institute of Cosmology and Gravitation, University of Portsmouth, Portsmouth, PO1 3FX, UK}

\author[0000-0003-4726-6714]{Gong-Bo Zhao}
\affiliation{National Astronomy Observatories, Chinese Academy of Science, Beijing, 100101, P.R.China}
\affiliation{University of Chinese Academy of Sciences, Beijing, 100049, P.R.China}
\affiliation{Institute of Cosmology and Gravitation, University of Portsmouth, Portsmouth, PO1 3FX, UK}

\author{Alex Zucca}
\affiliation{Department of Physics, Simon Fraser University, Burnaby, British Columbia, Canada V5A 1S6}



\begin{abstract}
We reconstruct evolution of the dark energy (DE) density using a nonparametric Bayesian approach from a combination of latest observational data. We caution against parameterizing DE in terms of its equation of state as it can be singular in modified gravity models, and using it introduces a bias preventing negative effective DE densities. We find a $3.7\sigma$ preference for an evolving effective DE density with interesting features. For example, it oscillates around the $\Lambda$CDM prediction at $z\lesssim0.7$, and could be negative at $z\gtrsim2.3$; dark energy can be pressure-less at multiple redshifts, and a short period of cosmic deceleration is allowed at $0.1 \lesssim z\lesssim 0.2$. We perform the reconstruction for several choices of the prior, as well as a evidence-weighted reconstruction. We find that some of the dynamical features, such as the oscillatory behaviour of the DE density, are supported by the Bayesian evidence, which is a first detection of a dynamical DE with a positive Bayesian evidence. The evidence-weighted reconstruction prefers a dynamical DE at a $(2.5\pm0.06)\sigma$ significance level.
\end{abstract}

\keywords{Cosmology: dark energy}

\section{Introduction} \label{sec:intro}
Since becoming the working model of cosmology following the discovery of cosmic acceleration \citep{Riess:1998cb, Perlmutter}, the $\Lambda$ Cold Dark Matter ($\Lambda$CDM) model withstood all the tests against increasingly accurate and comprehensive cosmic microwave background (CMB) \citep{WMAP,Ade:2015xua}, supernovae (SNe) \citep{Conley:2011ku,Suzuki:2011hu}, galaxy clustering \citep{SDSS} and weak lensing \citep{Heymans:2013fya,Hildebrandt:2016iqg} data. There are, however, good reasons to keep an open mind regarding possible extensions and alternatives to $\Lambda$CDM. The two main ingredients of the model, $\Lambda$ and CDM, are not understood at the fundamental level. Direct searches of dark matter have so far failed, while the observed value of the vacuum energy implies a technically unnatural fine-tuning of $\Lambda$ in the context of the effective field theory \citep{Burgess:2017ytm}, which is the framework for the present understanding of particle interactions.

The recent exquisite measurements of the CMB temperature and polarization by Planck \citep{Ade:2015xua} significantly reduced the uncertainties in $\Lambda$CDM parameters. With this dramatic improvement in precision, it is perhaps not surprising that several 2-3$\,\sigma$ level tensions have appeared between Planck and other datasets, as well as within the Planck data itself \citep{Addison:2015wyg}, when interpreted within the $\Lambda$CDM model. For instance, the locally measured value of the Hubble constant $H_0$ is off by $3.5\,\sigma$ from the Planck best fit \citep{Riess:2016jrr}. The expansion rate at $z=2.34$, implied by the Baryon Oscillation Spectroscopic Survey (BOSS) baryonic acoustic oscillations (BAO) measurement from the Lyman-$\alpha$ forest \citep{Font-Ribera:2013wce, Delubac:2014aqe}, disagrees with the best fit $\Lambda$CDM prediction at a $\sim2.7\,\sigma$ level. These tensions are not at a significance level sufficient to rule out $\Lambda$CDM -- they could simply be statistical fluctuations \citep{Scott:2018adl}. It is also possible that they are caused by unaccounted systematic effects in the measurements or the modelling of the data. However, it is worth noting that these tensions have persisted and got stronger over the past three years, fuelling significant interest in possible extensions of $\Lambda$CDM, such as dynamical dark energy (DE) \citep{Zhao:2017cud,runvac,ddetension,Sola:2017jbl,Ooba:2018dzf,Capozziello:2018jya}, interacting DE and dark matter \citep{Das:2005yj,Wang:2015wga,Yang:2018euj,idetension}, and other extensions of $\Lambda$CDM.

A reconstruction of the effective DE equation of state (EOS) $w_{\rm DE}^{\rm eff}(a)$ from a combination of recent data was presented in \citep{Zhao:2017cud}, showing a clear preference for $w_{\rm DE}^{\rm eff}<-1$, along with a possible crossing of the $w_{\rm DE}^{\rm eff}=-1$ divide. Such behaviour is impossible for quintessence -- a minimally coupled scalar field with a canonical kinetic term \citep{Ratra:1987rm,Wetterich:1994bg,Caldwell:1997ii,Steinhardt:1999nw,Wetterich:2001jw,PR02}, unless the sign of the kinetic term is changed by hand \citep{Caldwell:1999ew,Carroll:2003st,Vikman:2004dc}. This, however, leads to ghost instabilities, \ie~ energy being unbounded from below, unless one postulates a non-Lorentz-invariant cutoff at an appropriately chosen scale \citep{Cline:2003gs}. On the other hand, such ``ghostly'' behaviour generically occurs in non-minimally coupled DE models, \ie, where DE and matter interact through an additional scalar force \citep{Carroll:2004hc,Das:2005yj}. There, the $w_{\rm DE}^{\rm eff}$ measured in \citep{Zhao:2017cud} would not be the EOS of the scalar field, but it would be an effective quantity that depends on the coupling to matter. Such a $w_{\rm DE}^{\rm eff}$ is allowed to be phantom and to cross $-1$, with the effective dark energy density (defined below) allowed to change sign. In such cases, parametrizing the DE evolution via $w_{\rm DE}^{\rm eff}$ is unnecessarily restrictive as, by design, it does not allow for negative effective DE densities. To address this, we will directly reconstruct the effective DE density. 

We define the effective DE in a purely phenomenological way, by letting it describe all the contributions to the standard Friedmann equation other than matter and radiation. Namely, the effective DE density $\rho^{\rm eff}_{\rm DE}$ is defined via
\ba
{H^2 \over H^2_0}&=& {\Omega_{\rm r} \over a^{4}} + {\Omega_{\rm M} \over a^{3}} + \Omega_{\rm DE} X(a)\,,
\label{eq:def_rhoeff}
\ea
where $a$ is the scale factor, and $X(a) \equiv \rho^{\rm eff}_{\rm DE} (a) / \rho^{\rm eff}_{\rm DE} (1)$. $\rho^{\rm eff}_{\rm DE}$ could include modifications of gravity, non-minimal interactions with matter, \etc~ We assume a flat universe, so that $\Omega_{\rm r} + \Omega_{\rm M} + \Omega_{\rm DE}=1$ and $H(a=1)=H_0$. 

Most studies of DE dynamics (but not all, see \eg~\citep{WT04,Sahni:2006pa,Sahni:2014ooa,Poulin:2018zxs}) attempt to measure the DE EOS, $w_{\rm DE}=p_{\rm DE}/\rho_{\rm DE}$. If DE is a conserved fluid with $\rho_{\rm DE}>0$, then specifying $w_{\rm DE}(a)$ fully determines the dynamics of DE. In such a case, one can replace $p_{\rm DE}$ with $w_{\rm DE} \rho_{\rm DE}$ in the conservation equation, ${\dot \rho}_{\rm DE} + 3H \rho_{\rm DE}(1+ w_{\rm DE}) = 0$, and solve it to find $\rho_{\rm DE} (a) = \rho_{\rm DE}(1) \exp\left[\int_a^1 da' 3(1+w_{\rm DE}(a'))/a' \right]$. Working with the EOS provides a simple test of the $\Lambda$CDM as $w_\Lambda=-1$ independent of the value of $\Lambda$. However, in theories where the DE field mediates a force between matter particles, the effective DE EOS, $w^{\rm eff}_{\rm DE} \equiv p^{\rm eff}_{\rm DE}  / \rho^{\rm eff}_{\rm DE}$, can become singular, since $\rho^{\rm eff}_{\rm DE}$ can change sign (see the Appendix \ref{sec:appA} for details) \citep{Horndeski:1974wa,Deffayet:2009wt,Deffayet:2009mn,Deffayet:2011gz,Gleyzes:2014dya}. Thus, parameterizing the expansion history $w^{\rm eff}_{\rm DE}$ in cosmological codes used for model-independent tests of gravity, such as {\tt MGCAMB} \citep{Zhao:2008bn, Hojjati:2011ix}, {\tt EFTCAMB} \citep{Hu:2013twa, Raveri:2014cka} and {\tt hiCLASS} \citep{Zumalacarregui:2016pph}, could lead to a bias, as it assumes $\rho^{\rm eff}_{\rm DE}>0$ at all times. This is particularly relevant because the data indicate a preference for $w_{\rm DE}<-1$, which is prohibited for quintessence but can happen in modified gravity and brane-world models \citep{Damour:1990tw,Sahni:2002dx,Torres:2002pe,Chung:2002xj,Faraoni:2003jh,Carroll:2004hc,Das:2005yj}. For these reasons, looking directly at the evolution of DE density is more appropriate. Avoiding the assumption of a positive $\rho_{\rm DE}$ also allows one to constrain models such as ``Everpresent $\Lambda$'' \citep{Ahmed:2002mj,Ahmed:2012ci,Zwane:2017xbg} in which the observed cosmological ``constant'' fluctuates between positive and negative values.

\section{Data and Method}  
In what follows, we use the correlated prior method of  \citep{Crittenden:2011Dec, Crittenden:2009Dec} to reconstruct the evolution of the DE density from the available datasets used to probe the background expansion. We start with a brief review of the datasets and the reconstruction method.
 
Our phenomenological definition of the effective DE (\ref{eq:def_rhoeff}) does not specify the underlying theory needed for calculating cosmological perturbations. To keep the analysis general, we only evaluate observables predicted by the Friedmann equation and consider datasets probing the background expansion history. They include the CMB distance information from Planck \citep{Ade:2015xua}\footnote{The CMB distance prior used in this work was derived in \citep{Wang:2015tua} from the Planck2015 data, which is largely consistent with the latest Planck2018 result \citep{PLC18}.}, the ``Joint Light-curve Analysis" (JLA) supernovae \citep{Betoule:2014frx}, BAO measurements from 6dF Galaxy Survey (6dFGS) \citep{6df}, SDSS DR7 Main Galaxy Sample (MGS)\citep{MGS},  tomographic BOSS DR12 (TomoBAO) \citep{BAOWang}, eBOSS DR14 quasar sample (DR14Q) \citep{Ata:2017dya} and the Lyman-$\alpha$ forest (Ly$\alpha$FBAO) of BOSS DR11 quasars \citep{Font-Ribera:2013wce, Delubac:2014aqe}, the local measurement of $H_0=73.24 \pm 1.74 \rm \,[km\,s^{-1}\,Mpc^{-1}]$ in \citep{Riess:2016jrr}, and the Observational Hubble parameter Data (OHD) \citep{Moresco:2016mzx}.

\begin{figure}[t!]  
\centering
\includegraphics[width=0.5\textwidth]{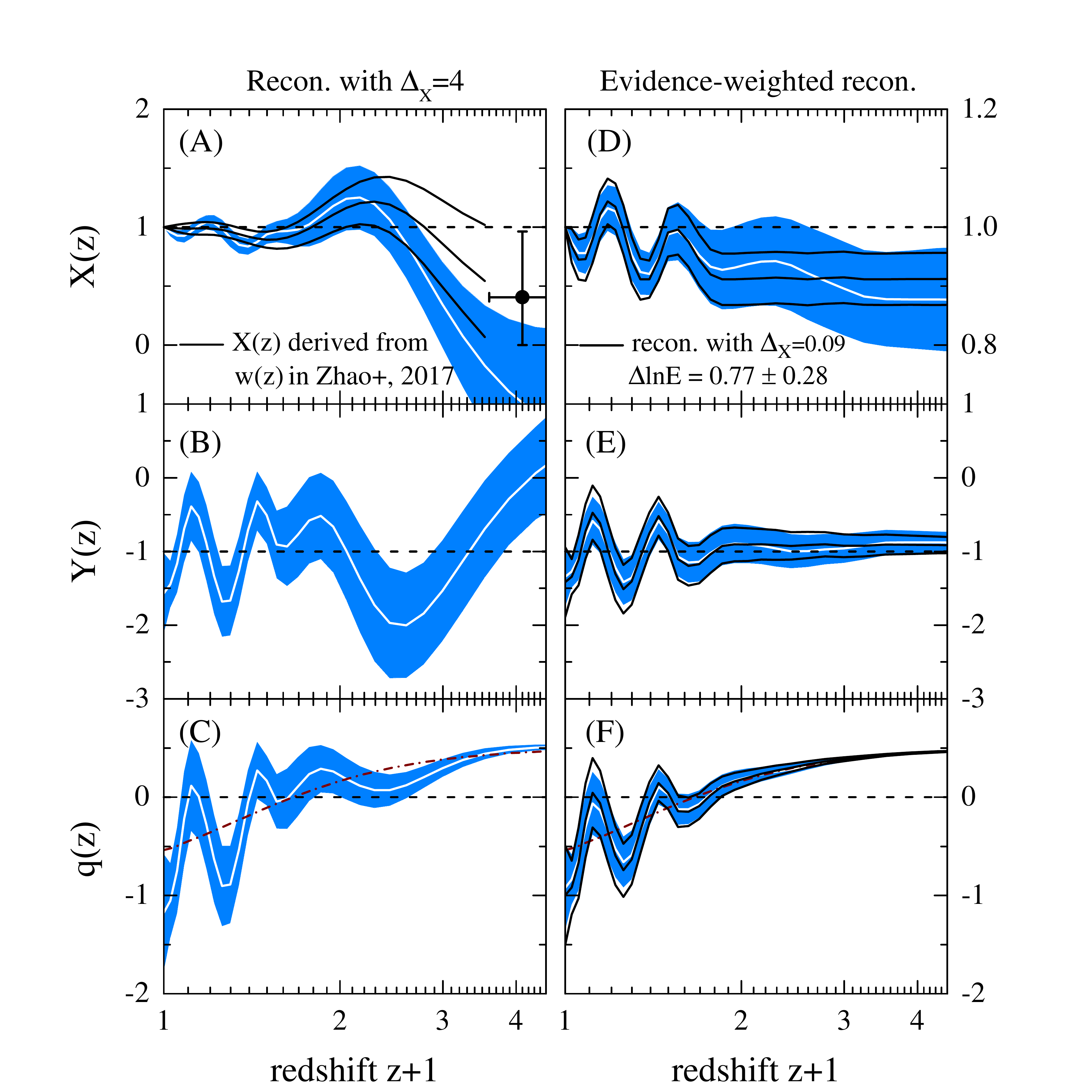}
\caption{Panel (A): $X(z)$ (best-fit and 68\% CL uncertainty) reconstructed using our standard correlated prior (blue filled band) compared to $X(z)$ derived from $w^{\rm eff}_{\rm DE}(z)$ reconstructed in \citep{Zhao:2017cud} (black curves and a data point with error bars); Panel (B): the reconstructed effective DE pressure $Y(z)\equiv P^{\rm eff}_{\rm DE}(z)/\rho^{\rm eff}_{\rm DE}(0)$; Panel (C): the reconstructed deceleration parameter $q(z)$. For reconstructions in panels (A-C), the range of variation of $X$ in each bin is set by $\Delta_X=4$. Panels (D-F): the same quantities as in panels (A-C) but reconstructed using $\Delta_X=0.09$ (black solid curves), and the evidence-weighted reconstruction defined in Eq.~(\ref{eq:weighted}) (blue filled band). The wine dash-dotted curves in panels (C) and (F) show the best-fit $q(z)$ in $\Lambda$CDM, and the dashed horizontal lines show $q=0$ to guide the eye. The dashed horizontal lines in panels (A,B,D,E) correspond to the $\Lambda$CDM model.} \label{fig:XYq}
\end{figure} 

To reconstruct $X(a)$, we parametrize it in terms of its values at $N$ points in $a$. Namely, we have bins $X_i=X(a_i)$, $i=1,...,N$, with $a_i$ distributed uniformly in the interval $a\in[1,0.001]$. We take $N=40$, which is large enough so that further increases do not affect the reconstruction results. If $X_i$ were assumed to be independent, fitting them to data would yield large uncertainties, rendering the reconstruction useless. Moreover, treating the bins as completely independent is an unreasonably strong assumption as, in any specific theory, the effective DE density would be correlated at nearby points in $a$. Motivated by these considerations, we use the method of \citep{Crittenden:2011Dec, Crittenden:2009Dec} and introduce a prior that correlates the neighbouring bins. Specifically, we take $X(a)$ to be a Gaussian random variable with a given correlation $\xi$ between its values at $a$ and $a'$, \ie, $\xi (|a - a'|) \equiv \left\langle [X(a) - X^{\rm fid}(a)][X(a') - X^{\rm fid}(a')] \right\rangle$. Here,  $X^{\rm fid}(a)$ is a reference fiducial model, and the correlation function $\xi$ is chosen so that it is nonzero for $|a-a'|$ below a given ``correlation length'' $a_c$, and approaches zero at larger separations. We adopt the CPZ form \citep{Crittenden:2011Dec, Crittenden:2009Dec} for the correlation, namely, $\xi(|a - a'|)=\xi(0)/[1+(|a - a'|/a_c)^2]$, where $\xi(0)$ sets the strength of the prior. The latter can be related to the expected variance in the mean value of $X$ as $\sigma^2_m \simeq \pi \xi(0){a_c}/(a_{\rm max}-a_{\rm min})$. In practice, we set $\sigma_m$ and $a_c$, and derive the corresponding $\xi(0)$. 

As our ``standard'' working prior we adopt $a_c=0.06$ and $\sigma_m=0.04$, which were the values used in \citep{Zhao:2017cud} to reconstruct $w_{\rm DE}(a)$. Their physical meaning is, of course, different, as $w_{\rm DE}$ and $X$ are related through an integral and, therefore, have different correlation properties. To better understand the impact of the prior, we have also performed reconstructions using different values of $a_c$ and $\sigma_m$. We find that our standard prior is rather weak in the sense that decreasing $a_c$ or increasing $\sigma_m$ increases the uncertainties but does not change the shape of the reconstruction. We show results with a few stronger priors in the Appendix \ref{sec:appB}. As expected, in the limit of $a_c \rightarrow 1$ or $\sigma_m \rightarrow 0$ the reconstructed $X$ becomes a constant.

Our binning scheme for $X$ includes eight bins in the redshift range $3<z<1000$, between the redshift of the Ly$\alpha$FBAO measurement and the CMB last scattering surface. The reason for keeping these eight bins, even though there are no data points in that range, was to simplify the evaluation of the prior covariance which can be performed analytically for uniformly spaced bins. However, physically, no information can be gained from having these additional degrees of freedom at $z>3$. To check the role they play in the reconstruction, we replaced them with a single bin for $z\in[3.85,1000]$. The value of $X$ in the wide bin was either fixed to $1$, or allowed to vary freely, uncorrelated with the 32 lower redshift bins. As shown in the Appendix \ref{sec:appB} , the reconstruction of $X$ in the range probed by observations, $z\in[0, 3]$, is largely unaffected by what is assumed at higher redshifts.
 
\begin{figure}[t!]  
\centering
\includegraphics[scale=0.27]{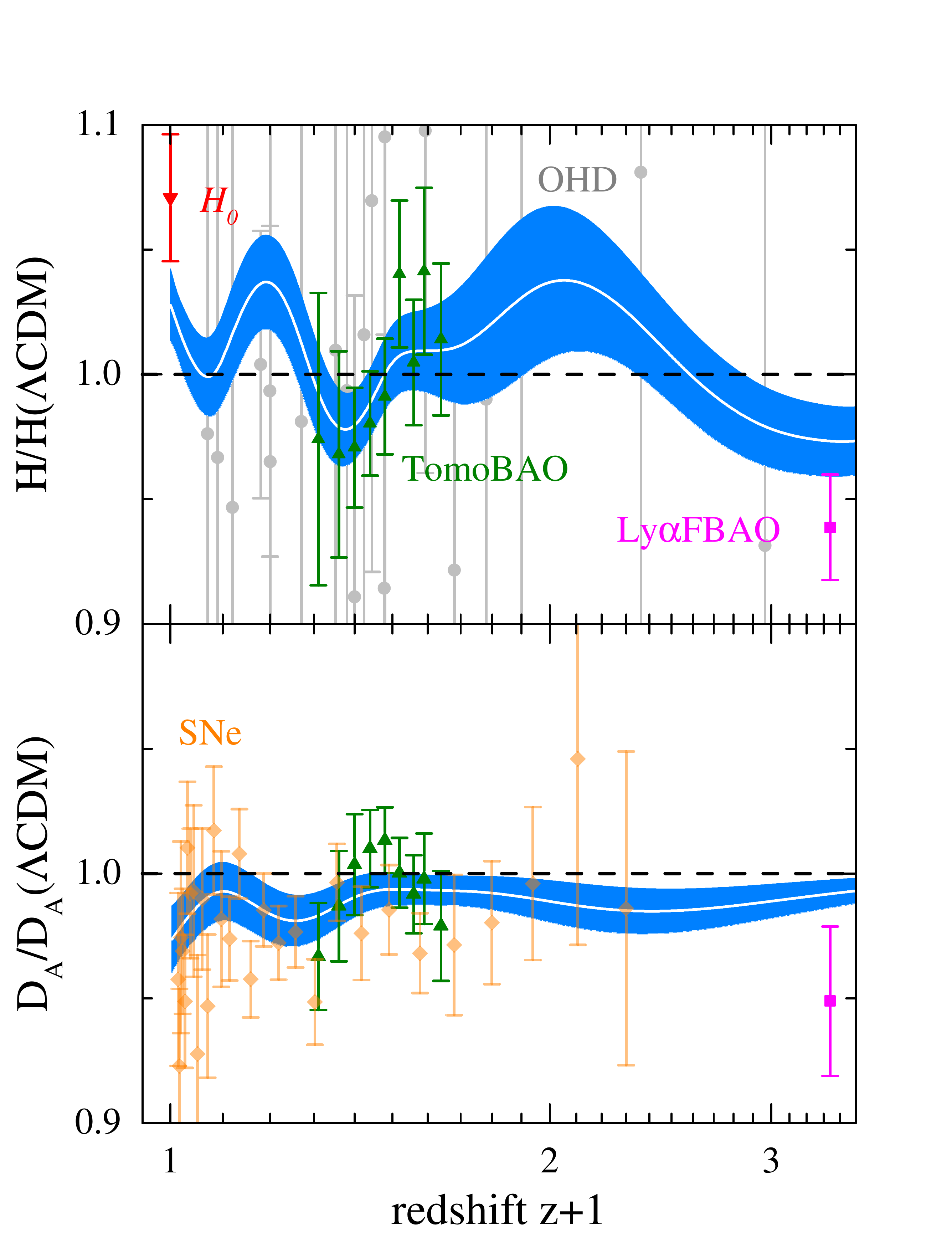}
\caption{Top panel: $H(z)$ derived from the reconstructed $X(z)$ rescaled by that of the best fit $\Lambda$CDM model. The data points with error bars show the measurements of $H$ as illustrated in the legend; Lower panel: same as the top panel but for the angular diameter distance $D_A$.} 
\label{fig:Xdata}
\end{figure} 
 
We use {\tt PolyChord}\footnote{\url{https://ccpforge.cse.rl.ac.uk/gf/project/polychord/}} \citep{polychordI,polychordII}, a nested sampling plug-in for {\tt CosmoMC}\footnote{\url{http://cosmologist.info/cosmomc/}} \citep{cosmomc} which enables computation of the Bayesian evidence, to sample the parameter space $\mathbf{P}=\{\Omega_bh^2, \Omega_ch^2, \Theta_s, X_i, \mathcal{N}\}$, where $\Omega_{b}h^2$ and $\Omega_{c}h^2$ are the physical densities of baryon and CDM respectively, $\Theta_s$ is the ratio of the sound horizon to the angular diameter distance at the decoupling epoch (multiplied by $100$), $X_i~(i=1,...,40)$ are the binned DE density parameters, and $\mathcal{N}$ collectively denotes all the data nuisance parameters that need to be marginalised over. All parameters are sampled from sufficiently wide flat priors. In particular, the range for the $X$ bins is set to be $[1-\Delta_X,1+\Delta_X]$ with $\Delta_X=4$ as the default value. Note that the value of $X$ in the first bin is fixed by definition in Eq.~(\ref{eq:def_rhoeff}): $X_1 = X(a=1)=1$. The range of nearby bins is then effectively reduced by the correlations induced by the prior. In addition, the sampling procedure guarantees $H^2(a)>0$ at all times. To test how the reconstruction and Bayesian evidence for dynamical DE change with the strength of the prior, we additionally perform reconstructions for a set of $\Delta_X$ values logarithmically spaced in the interval $\Delta_X\in[0.01,4]$. We also run a special case with $\Delta_X=0$, which represents the $\Lambda$CDM model, for a comparison. A modified version of {\tt CAMB}\footnote{\url{http://camb.info}} \citep{CAMB} is used to calculate theoretical predictions. 

\begin{table} [b!]
\scriptsize
\caption{The upper part: the difference in $\chi^2$ between $X$CDM and $\Lambda$CDM models, $\Delta \chi^2 \equiv \chi^2_{\rm XCDM} -  \chi^2_{\rm \Lambda CDM}$, for individual datasets; the lower part: the significance, $\rm S/N \equiv \sqrt{|\Delta \chi^2|}$ and the Bayesian evidence $\rm \Delta\,ln E$ for reconstructions using $\Delta_X=4$ and $\Delta_X=0.09$, and the evidence-weighted reconstruction.} \label{tab:chisq}
\begin{center}
\begin{tabular}{c|c|c|c|c|c|c}
\hline\hline
				&	TomoBAO		&	Ly$\alpha$FBAO		&	SNe	       &	$H_0$		&	OHD	 & prior	\\
\hline
$\Delta \chi^2$	       &	$-4.9$	       &	$-4.3$			       &	$-4.1$	&	$-4.1$	       &	$-1.2$	& $+3.0$\\
\hline
				&	\multicolumn{6}{c}{\rm All data + standard correlated prior}		\\
				\cline{2-7}
				&      \multicolumn{2}{c|}{$\Delta_X=4$}    &  \multicolumn{2}{c|}{$\Delta_X=0.09$}   &  \multicolumn{2}{c}{evidence-weighted} \\
\hline
S/N			        &	\multicolumn{2}{c|}{$3.7\sigma$}     & \multicolumn{2}{c|}{$2.8\sigma$}	&   \multicolumn{2}{c}{$(2.5\pm0.06)\sigma$} \\	
$\rm \Delta\,ln E$      &	\multicolumn{2}{c|}{$-5.7\pm0.3$}   & \multicolumn{2}{c|}{$0.77\pm0.28$}	& \multicolumn{2}{c}{N/A} \\	
\hline\hline                                        
\end{tabular}
\end{center}
\end{table}

\section{Results} 
Panel (A) of Fig. \ref{fig:XYq} compares our reconstructed $X(z)$ to the one derived from $w^{\rm eff}_{\rm DE}(z)$ reconstructed in \citep{Zhao:2017cud} using similar data, showing them being mutually consistent. The $X(z)$ derived from $w^{\rm eff}_{\rm DE}(z)$ is strictly positive as using the EOS implicitly assumes the positivity of the DE density that can bias the reconstruction. Separate reconstructions from individual datasets are presented in the Appendix \ref{sec:appC} , showing that the $H_0$ and the Ly$\alpha$FBAO data both drag $X(z)$ towards negative values at high $z$. The best-fit parameters of XCDM are given in Table \ref{tab:paras}. From the values of $X$ at various redshifts, we can see that the oscillatory features of energy density is favored by the combined data.

\begin{table}
\caption{The best-fit values and the 68\% CL uncertainties of cosmological parameters in the $\Lambda$CDM and $X$CDM models.} \label{tab:paras}
\begin{center}
\begin{tabular}{c|c|c}
\hline\hline
			         &$\Lambda{\rm CDM}$	        &$X{\rm CDM}$	                         \\    \hline
$\Omega_bh^2$	&$0.0224\pm0.00013$	        &$0.0223\pm0.00016$ 	        \\
$\Omega_ch^2$	&$0.1189\pm0.00087$	        &$0.1203\pm0.00141$		\\
$\Theta_s$		&$1.0412\pm0.00029$	        &$1.0411\pm0.00032$		\\
$X(z=0.08)$                  &$1$	        &$0.926\pm0.056$		\\
$X(z=0.18)$                  &$1$	        &$1.042\pm0.062$		\\
$X(z=0.39)$                  &$1$	        &$0.835\pm0.068$		\\
$X(z=1.16)$                  &$1$	        &$1.251\pm0.282$		\\
$X(z=2.24)$                  &$1$	        &$0.076\pm0.438$		\\
\hline
$\Omega_{\rm M}$	&$0.302\pm0.005$			&$0.288\pm0.008$		\\
$H_0 \rm [km s^{-1} Mpc^{-1}]$	 		&$68.41\pm0.387$		        &$70.30\pm0.998$\\
\hline\hline
\end{tabular}
\end{center}
\end{table}%

Fig.~\ref{fig:Xdata} compares $H(z)$ and $D_A(z)$ derived from the reconstructed $X(z)$ to their observed values, with all quantities rescaled by their best-fit $\Lambda$CDM values. The oscillatory features in the derived $H(z)$ and $D_A(z)$ at $z\lesssim0.7$, which are directly related to those in $X(z)$ shown in panel (A) of Fig. \ref{fig:XYq}, are driven by measurements of $H_0$, SNe and TomoBAO. The bump-and-damp feature in $H(z)$ at $z\gtrsim0.7$ (also seen in the reconstruction performed in \citep{Poulin:2018zxs}), on the other hand, is due to the Ly$\alpha$FBAO measurement and the integral constraint of the CMB. This can be read from the improved $\chi^2$ listed in Table \ref{tab:chisq}, namely, the $X$CDM model reduces the $\chi^2$ of TomoBAO, Ly$\alpha$FBAO, $H_0$ and SNe by $4.9$, $4.3$, $4.1$ and $4.1$ respectively, meaning that it is these datasets that contribute the most to the features in $X$.

In Fig. \ref{fig:XYq}, in addition to $X$, we show the normalized effective DE pressure, $Y(z)\equiv P^{\rm eff}_{\rm DE}(z)/\rho^{\rm eff}_{\rm DE}(0)$\footnote{We divide the effective pressure by the energy density today instead of $\rho^{\rm eff}_{\rm DE}(z)$ to avoid a singularity when $\rho^{\rm eff}_{\rm DE}(z)$ changes sign.} derived via $Y=-X+\frac{1}{3}{\rm d}X/{\rm d}z(1+z)$, and the deceleration parameter, $q(z)$. As shown in panel (B), $Y(z)$ oscillates around the $\Lambda$CDM prediction of $-1$, and, interestingly, DE is within $\sim1\sigma$ of having zero pressure at $z\simeq0.1$, $0.5$, $0.9$ and $z\gtrsim3$. Panel (C) shows $q(z)$ oscillating around the prediction of the best-fit $\Lambda$CDM model (dash-dotted line). Unlike in $\Lambda$CDM, where the acceleration starts at $z\sim 0.6$, the best fit $X$CDM universe would not be accelerating until $z\sim0.45$, and would experience a short period of deceleration during $0.1\lesssim z \lesssim0.2$, although it is far from conclusive given the uncertainties.

The best-fit dynamical DE model reduces the $\chi^2$ by $13.9$ compared to $\Lambda$CDM, which implies it being preferred at a 3.7$\,\sigma$ level. However, $X$CDM has more degrees of freedom, and the appropriate way to assess the significance of DE dynamics is to compare the Bayesian evidences. For our ``standard" prior, the Bayes factor, which is the logarithm of the ratio of the evidences, is negative, indicating no evidence for the best-fit $X$CDM.

\begin{figure}[!b]  
\includegraphics[scale=0.25]{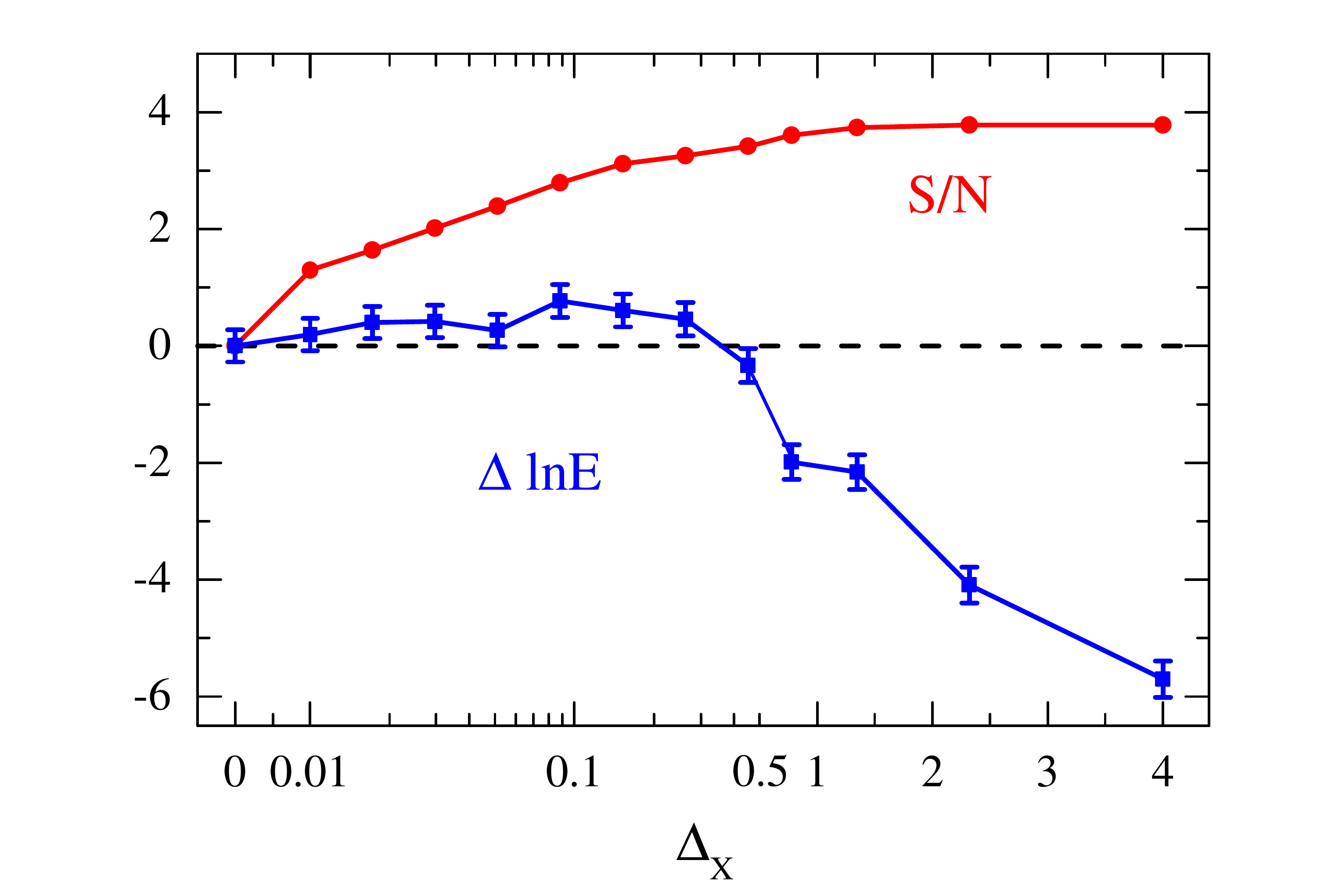}
\caption{The Bayes factor relative to that of the $\Lambda$CDM ($\Delta \,\rm lnE$) and the significance level of deviation from $\Lambda$CDM ($\rm S/N \equiv \sqrt{|\Delta \chi^2|}$) for different values of $\Delta_X$ that set the range of variation of $X$ in each bin.} \label{fig:chi2lnE}
\end{figure} 

To see the extent to which these conclusions depend on the choice of the prior parameters, we perform reconstructions with different prior strength and show, in Fig.~\ref{fig:chi2lnE}, the significance, $\rm S/N = \sqrt{|\Delta \chi^2|}$, and the evidence as functions of $\Delta_X$ (the range of variation of $X$ in each bin). We see that, as expected, both the S/N and the evidence approach zero when the prior is very strong. Interestingly, the evidence shows a trend of climbing towards positive values as $\Delta_X$ increases, with a peak showing up at $\Delta_X\sim0.1$, and drops below zero for $\Delta_X\gtrsim0.4$. This motivates us to consider an evidence-weighted reconstruction, which linearly combines reconstructions with different $\Delta_X$, weighted by the Bayesian evidence, \ie,
\ba
Z_W(z)&\equiv&\frac{\sum_i \left[Z(z;\Delta_{X_i}) e^{\Delta \ln{\rm E}(\Delta_{X_i})}\right]}{\sum_i \left[ e^{\Delta \ln{\rm E}(\Delta_{X_i})}\right]} \,
\label{eq:weighted}
\ea where $Z=X, Y, q$. The evidence-weighted reconstructions are shown in panels (D-F) of Fig. \ref{fig:XYq}. They retain the key features of the best-fit $X$CDM shown in panels (A-C), but at a lesser significance. In particular, the overall significance of the deviation from $\Lambda$CDM reduces to $(2.5\pm0.06)\sigma$. Panels (D-F) show the the results for $\Delta_X=0.09$, which corresponds to the maximal Bayesian evidence $\Delta{\rm lnE}=0.77\pm0.28$. We find them to be quite similar, as expected, since the linear combination in the evidence-weighted reconstruction is dominated by the component with maximal weight.

\section{Conclusion and Discussions} \label{sec:conclusion}
Different levels of tension among various kinds of observations within the framework of $\Lambda$CDM necessitates the exploration of extended cosmological models beyond $\Lambda$CDM. As was shown in an earlier study \citep{Zhao:2017cud}, dynamical DE parameterized in terms of its EOS is able to release the tension, but is not favoured over $\Lambda$CDM by the Bayesian evidence.

In this work, we take a another route to investigating the evolution of DE, namely, we directly reconstruct the effective DE density $X$ from data. This is advantageous over reconstructing the EOS, since $X$ is more directly related to data, while the EOS is related to the derivative of $H(z)$. Thus, we caution against parameterizing DE in terms of its EOS as it generally biases towards positive values and smoother evolution of the DE density. This can affect the Bayesian evidence -- one can obtain from the same data a positive Bayes while using $X(z)$, and a negative one when using $w_{\rm  DE}(z)$. Furthermore, $X$ is more physically relevant in modified gravity theories, where it can be negative and change sign. Such dynamics would be forbidden for $X$ derived from an EOS.

We find hints of DE dynamics at a significance of $3.7\sigma$ with interesting features. For example, $X$ oscillates around the $\Lambda$CDM prediction at $z\lesssim0.7$, and can become negative at $z\gtrsim2.3$; DE can be pressure-less at multiple redshifts during evolution, and a short period of cosmic deceleration is allowed by current data at $0.1\lesssim z\lesssim0.2$. We note that these features would have been missed if the DE density was parameterized using a simple polynomial \citep{LLEG18}. Some of these dynamical features, including the oscillations, are supported by the Bayesian evidence (the Bayesian factor is positive at about $2.8\sigma$ level for the case of $\Delta_X=0.09$ for example), which is the first time a dynamical DE with a positive Bayesian evidence is detected. Furthermore, the evidence-weighted reconstruction prefers the dynamical DE at a $(2.5\pm0.06)\sigma$ significance level.

The new features of DE dynamics await scrutiny by forthcoming BAO measurements by Dark Energy Spectroscopic Instrument (DESI)\footnote{\url{http://desi.lbl.gov/}}, Euclid\footnote{\url{https://www.euclid-ec.org}} and Prime Focus Spectrograph (PFS)\footnote{\url{http://pfs.ipmu.jp/}}. Gravitational wave sources accompanied by electromagnetic counterparts will also offer accurate independent estimates of $H_0$ at very low redshifts \citep{Abbott:2017xzu,Guidorzi:2017ogy,Hotokezaka:2018dfi}. The methodology developed in this work will be useful in further studies of DE and modified gravity.

\acknowledgements

We benefited from valuable discussions and previous collaborations with Rob Crittenden. We thank George Efstathiou, Wenjuan Fang and Marco Raveri for useful discussions. YW and  GBZ are supported by NSFC Grants 1171001024 and 11673025, and the National Key Basic Research and Development Program of China (No. 2018YFA0404503). YW is also supported by the Young Researcher Grant of NAOC. LP and AZ are supported by the National Sciences and Engineering Research Council of Canada. This research used resources of the SCIAMA cluster supported by University of Portsmouth, and the ZEN cluster supported by NAOC.

\appendix

\twocolumngrid

\section{The effective dark energy density in general cosmologies} \label{sec:appA}

In what follows, we show that in modified gravity theories the effective DE density $X(a)$, defined in Eq. (\ref{eq:def_rhoeff}),
is allowed to change sign. To see this, consider the class of generalized Brans-Dicke (GBD) models described by the action \citep{Bergmann1968, Nordtvedt1970, Wagoner1970}, 
\ba
\nonumber
S = \int d^4x \sqrt{-g} \Big[ {F(\phi) R \over 16 \pi G} - {1 \over 2}  \partial^\mu \phi \partial_\mu \phi 
- V(\phi) + {\cal L}_M \Big]\, ,
\ea
where ${\cal L}_M$ is the Lagrangian of all particle and radiation fields. The modified Einstein equation in this model is, 
\ba
\nonumber
G_{\mu \nu} &=& 8 \pi G F^{-1} \left\{ T^M_{\mu \nu} + T^\phi_{\mu \nu} + \nabla_\mu \nabla_\nu F- g_{\mu \nu} \Box F \right\} \\
&=& 8 \pi G  \left\{ T^M_{\mu \nu} + (T^{\rm eff}_{\rm DE})_{\mu \nu} \right\}\,,
\label{eq:modeinstein}
\ea
where, in the second line, we have defined the effective DE stress-energy by absorbing into it all the terms on the right hand side other than the usual matter term, {\ie},
\be
\nonumber
(T^{\rm eff}_{\rm DE})_{\mu \nu} \equiv F^{-1} \left\{ T^\phi_{\mu \nu} + \nabla_\mu \nabla_\nu F- g_{\mu \nu} \Box F + (1-F) T^M_{\mu \nu} \right\}\,.
\ee
Then, the effective DE density is 
\be
\rho^{\rm eff}_{\rm DE} = F^{-1} \left\{\dot{\phi}^2/2 + V(\phi) - 3H \dot{F} +(1-F)\rho_M \right\}\,,
\ee
while the effective DE pressure is
\be
p^{\rm eff}_{\rm DE} = F^{-1} \left\{\dot{\phi}^2/2  - V(\phi) +2H\dot{F}+\ddot{F} \right\}\,.
\ee
The $\mu=\nu=0$ component of Eq.~(\ref{eq:modeinstein}) for a Friedmann-Robertson-Walker background metric gives the usual Friedmann equation,
\be
H^2 = \left( \dot{a} \over a \right)^2 = {8 \pi G \over 3} [ \rho_M(a) + \rho^{\rm eff}_{\rm DE}(a)],
\ee
which can then be recast in the form of Eq.~(\ref{eq:def_rhoeff}).

The effective DE ``fluid'' is, by construction, conserved,
\be
{\dot \rho}^{\rm eff}_{\rm DE} + 3H (\rho^{\rm eff}_{\rm DE}+ p^{\rm eff}_{\rm DE})=0\,,
\ee
but its EOS,
\be
w^{\rm eff}_{\rm DE}= {\dot{\phi}^2/2 - V(\phi) +2H\dot{F}+\ddot{F} \over \dot{\phi}^2/2 + V(\phi)- 3H \dot{F} +(1-F)\rho_M}\,,
\ee
is not always well-defined because $\rho^{\rm eff}_{\rm DE}$ in the denominator is allowed to change sign thanks to the new terms generated by the non-minimal coupling $F(\phi)$. In the case of quintessence, $F=1$, the effective DE EOS is the same as the EOS of the scalar field, $w_\phi \ge -1$. For a general $F(\phi)$, the scalar field mediates a force between matter particles, coupling the matter fluid with DE so that they are no longer separately conserved. Thus, as articulated in \citep{Carroll:2004hc,Das:2005yj}, observing $w^{\rm eff}_{\rm DE}<-1$, or finding that $\rho^{\rm eff}_{\rm DE}$ changes its sign, could be a smoking gun for new interactions in the dark sector. 

\section{Investigating alternative priors and high-$z$ parameterizations of $X(z)$} \label{sec:appB}

\begin{figure}
\centering
\includegraphics[scale=0.3]{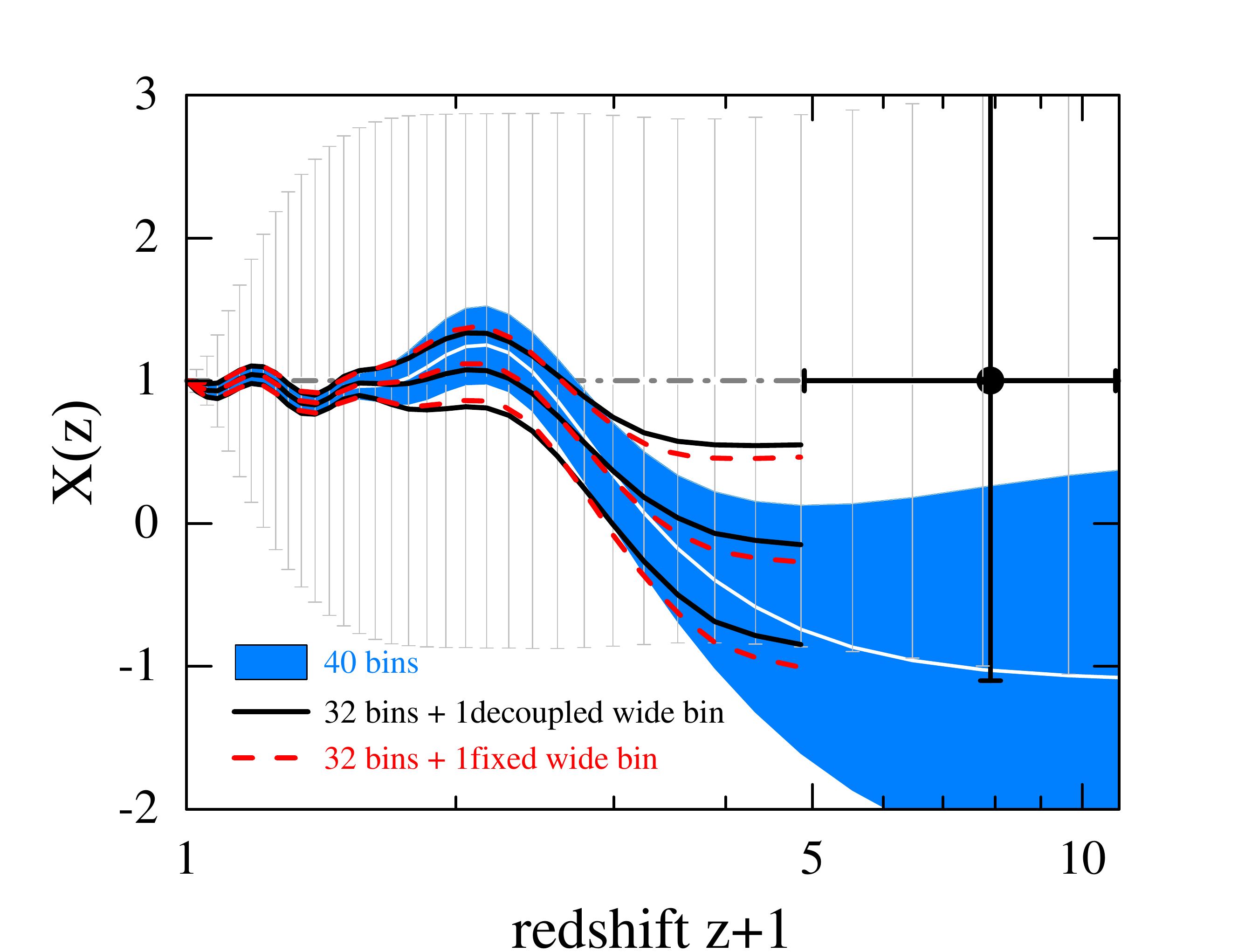}
\caption{The reconstructed evolution of $X(z) \equiv \rho^{\rm eff}_{\rm DE}(z)/\rho^{\rm eff}_{\rm DE}(0)$ (white line with the 1$\sigma$ blue band around it) obtained by fitting 40 bins uniformly spaced in $a\in[1,0.01]$ with the help of our standard prior ($a_c=0.06$, $\sigma_m=0.04$). The discrete error bars show the 1$\sigma$ uncertainties on the bins from the prior alone. This reconstruction is compared to two cases where the last 8 bins, in the $a\in[0.001,0.206]$ range ($3.85<z<1000$), are replaced with a single wide bin: one in which it is allowed to vary independently (red solid lines showing the best fit and the 1$\sigma$ band) and one where it is fixed to 1 (green dashed lines).} \label{fig:comlastbin}
\end{figure}

\begin{figure*} 
\centering
\includegraphics[scale=0.45]{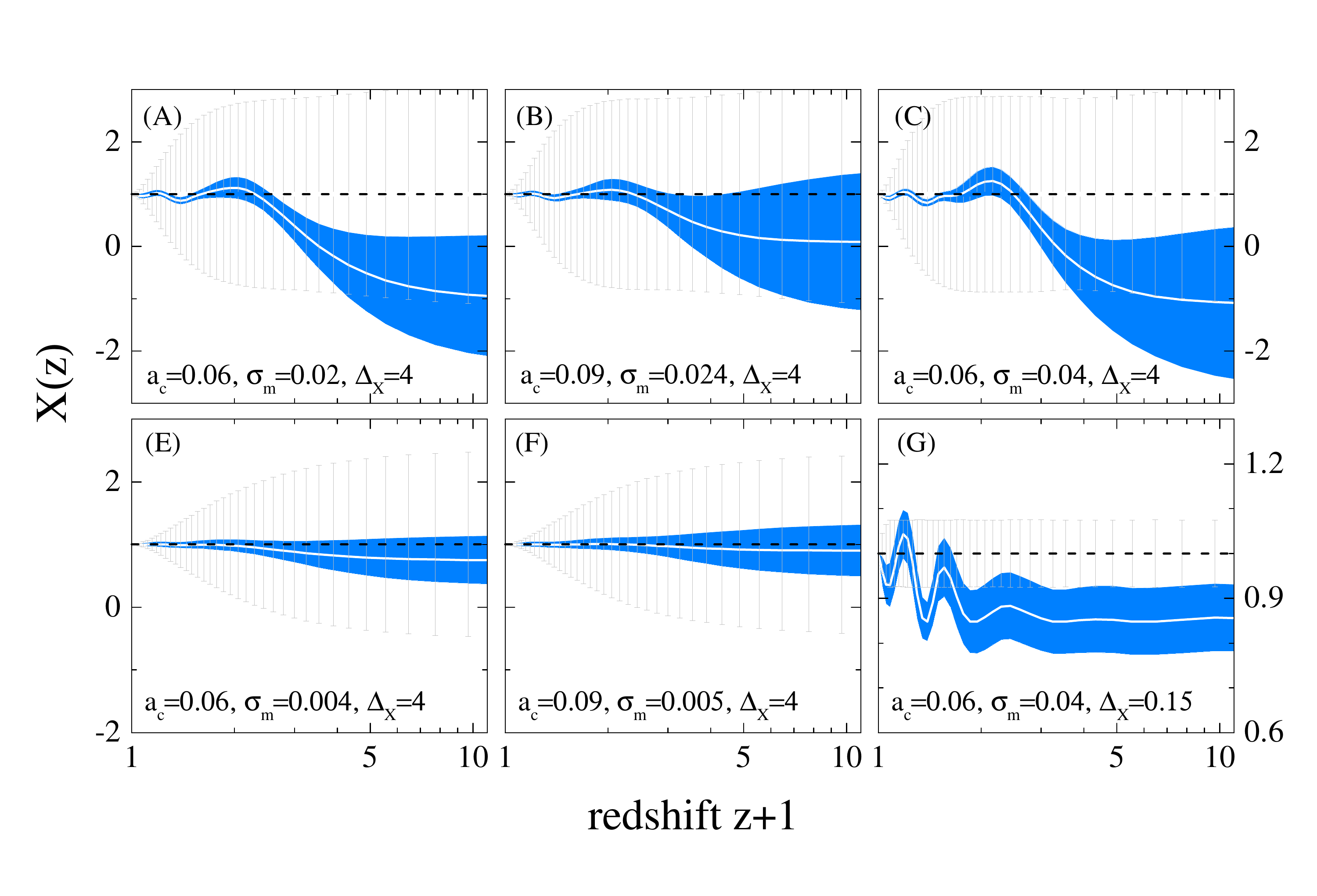}
\caption{Comparison of the $X(z)$-reconstructions obtained using different values of the correlated prior parameters $a_c$ and $\sigma_m$, and the range $\Delta_X$ over which $X$ can vary in each bin. The discrete error bars show the 1$\sigma$ uncertainties from the prior alone. Panel (C) is the case of our standard prior, also shown in Fig.~\ref{fig:comlastbin}. Note that the vertical axis range in Panel (G) differs from that in other panels.} \label{fig:diffpriors}
\end{figure*} 

The reconstructed evolution of $X(z)$ from the combination of all data using our ``standard'' prior is shown in Fig.~\ref{fig:comlastbin}. To help interpret the reconstruction, we also show the 1$\sigma$ uncertainties around on the $40$ bins from our Gaussian prior alone. The latter are obtained by running {\tt CosmoMC} and letting it converge with using just the prior and no data. The fiducial model assumed by the prior is $X(z)=1$, and the ``best fit'' to the prior alone is very close, although not identical to it, as expected. One can see that at lower redshifts the data significantly improves on the prior, while at high redshifts the reduction in uncertainties is relatively small.

There are no data points probing the expansion history between $z=2.34$, where the Lyman-$\alpha$ forest provides a BAO measurement, and the epoch of last scattering at $z \sim 1000$ probed by CMB. Thus, having several bins of $X(z)$ in that redshift range is not justified, except for the purpose of keeping the spacing between the bins uniform. To check that these ``extra''  bins do not affect the reconstruction of $X(z)$, we try a couple of alternative choices of parametrizing $X$ at $z > 2.34$. Specifically, we tried replacing the last 8 bins in the $a\in[0.001,0.206]$ range ($3.85<z<1000$) with a single wide bin and either fixed it to $X=1$ or let it vary independently from other bins. As shown in Fig.~\ref{fig:comlastbin}, the reconstructed dynamics and the size of uncertainties at $z<3$ remains consistent in all three cases.

Finally, in Fig.~\ref{fig:diffpriors}, we show the effect of using alternative prior parameters in our reconstructions. There are two types of prior parameters: ($a_c$, $\sigma_m$) that set the prior covariance of the bins, and $\Delta_X$ which sets the range of allowed values of $X$ in each bin. In each case, we also show the 1$\sigma$ uncertainties on the bins from the prior alone. The significance of the DE dynamics detection in Panels (A) and (B) is $3.4\,\sigma$ and $2.9\,\sigma$, respectively. For the stronger priors, \ie, Panels (E) and (F), the reconstructions are consistent with $\Lambda$CDM.

\section{The $X(z)$ reconstructions using different data combinations} \label{sec:appC}

\begin{figure*}
\centering
\includegraphics[scale=0.45]{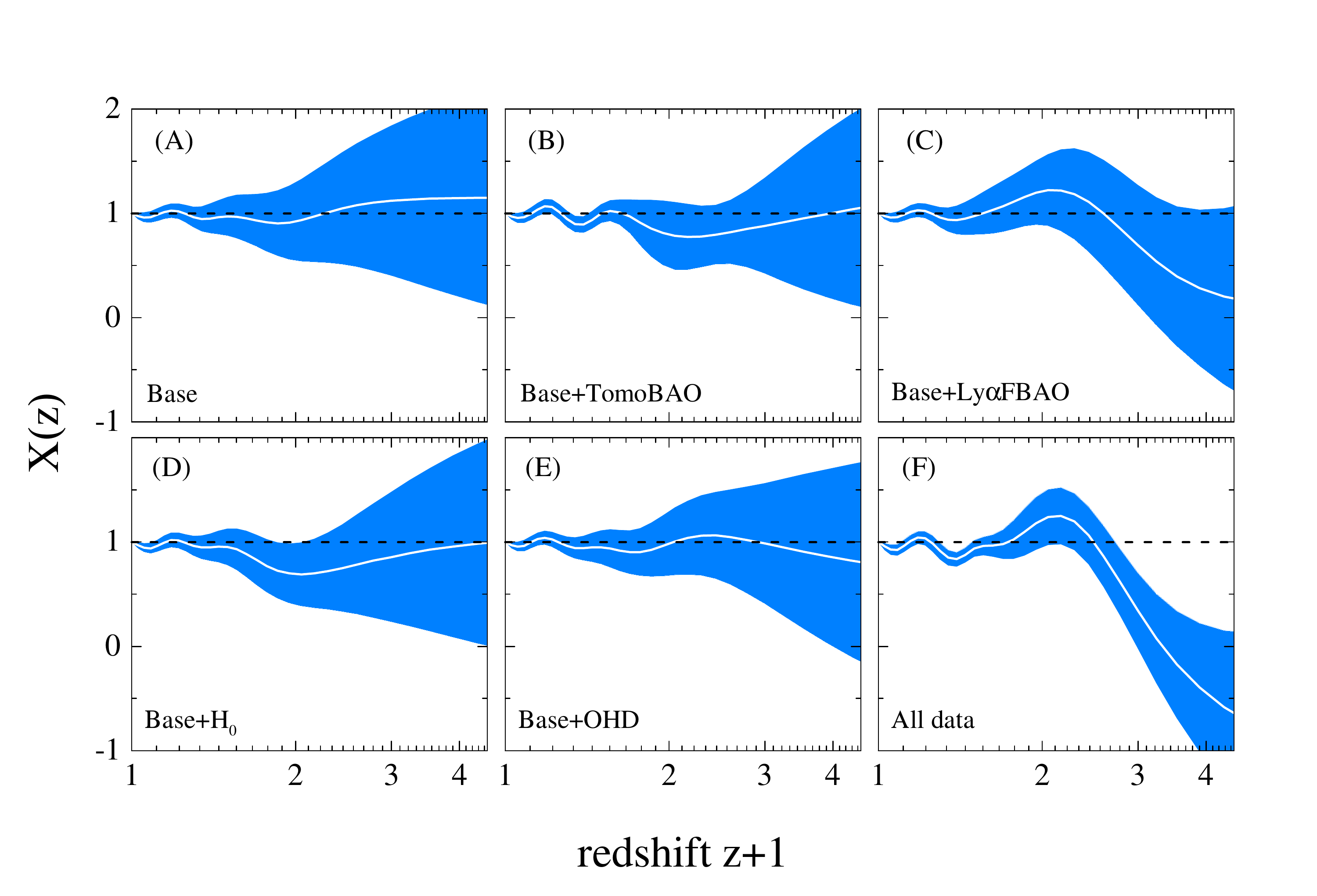}
\caption{Panels (A-F): The best-fit and the 68\% CL uncertainty of $X(z)$ reconstructed from six different data combinations.} \label{fig:diffdata}
\end{figure*}

To show how each particular dataset affects the reconstruction result, we reconstruct $X(z)$ from various data combinations. We first combine a collection of datasets with no reported tensions among themselves into a ``Base" dataset, which includes the 2015 Planck distance priors, JLA supernovae, and BAO measurements from 6dFGS, MGS and eBOSS DR14Q \footnote{{This work additionally uses the eBOSS BAO data, while excluding the galaxy power spectra, redshift space distortions, weak lensing and CMB anisotropies which require modelling of perturbations and were used in the EOS reconstruction \citep{Zhao:2017cud}}.}. We then add other datasets, one at a time, to ``Base" to form other data combinations, finally combining all data together to form the most constraining dataset. The reconstructions of $X(z)$, \ie~the best fit values and the 68\% confidence level (CL) uncertainties for each bin, from these datasets are presented in panels (A-F) of Fig.~\ref{fig:diffdata}. One can see that $X(z)$ reconstructed from either Base or Base+OHD is consistent with that predicted by $\Lambda$CDM, while the reconstruction derived from the other four data combinations show different levels of dynamics in $X(z)$. For example, results derived either from Base+$H_0$ or Base+Ly$\alpha$FBAO prefer a lower $X(z)$ at higher redshifts. Specifically, Base+Ly$\alpha$FBAO diminishes $X$ at $z\sim2.3$ (it even makes $X$ negative at $z\gtrsim2.3$), which is due to the fact that the Ly$\alpha$FBAO measurement at $z\sim2.3$ is lower than the theoretical prediction of $\Lambda$CDM at around $2.5\sigma$ \citep{Font-Ribera:2013wce, Delubac:2014aqe}. On the other hand, Base+$H_0$ drags $X$ downwards at $z\sim1$. This is because the local measurement of $H_0$ prefers a much greater value than that extrapolated from the best fit $\Lambda$CDM. The most effective way to fit a higher $H_0$ is to increase $X(z=0)$, but as $X(z=0)$ is fixed to unity by definition, $X(z)$ has to be reduced at higher redshifts, namely, at $z\simeq1$\footnote{$X$ is not allowed by SNe, which have a strong constraining power, to deviate from unity at $z\lesssim1$. On the other hand, DE becomes dynamically unimportant at $z \gg 1$. Thus, to compensate for a higher $H_0$, $X$ is reduced at $z\simeq1$.}. Note that, these redshift-dependent reductions of $X$ caused by $H_0$ and Ly$\alpha$FBAO would be degenerate if one fit a constant $X$ instead, as both datasets ``pull'' $X$ in the same direction. When the tomographic BAO is added to Base, a statistically significant oscillatory feature shows up at $z\lesssim 0.6$ (see panel B). Specially, $X(z)$ tends to go below unity at $z\sim 0.1$ and $z\sim 0.3$ and above unity at $z\sim 0.2$. Such details would not be revealed without the high redshift resolution BAO measurements \citep{BAOWang, BAOZhao}. The same oscillatory feature at $z\lesssim 0.6$ is present in the SNe data, and becomes more significant when all data are combined, as shown in panel (F). The decrease in $X$ at $z\gtrsim 1.5$ also becomes more pronounced. In addition, a new bump appears at $z\sim 1.3$ caused by the requirement to maintain a fixed distance to last scattering, set by CMB measurements, while compensating for the reduction in $X$ at high redshifts.

\begin{thebibliography}{}

\bibitem[Riess et al.(1998)]{Riess:1998cb} Riess, A.~G., Filippenko, A.~V., Challis, P., et al.\ 1998, \aj, 116, 1009 

\bibitem[Perlmutter et al.(1999)]{Perlmutter} Perlmutter, S., Aldering, G., Goldhaber, G., et al.\ 1999, \apj, 517, 565

\bibitem[Bennett et al.(2013)]{WMAP} Bennett, C.~L., Larson, D., Weiland, J.~L., et al.\ 2013, \apjs, 208, 20 

\bibitem[Planck Collaboration et al.(2016)]{Ade:2015xua} Planck Collaboration, Ade, P.~A.~R., Aghanim, N., et al.\ 2016, \aap, 594, A13 

\bibitem[Conley et al.(2011)]{Conley:2011ku} Conley, A., Guy, J., Sullivan, M., et al.\ 2011, \apjs, 192, 1 

\bibitem[Suzuki et al.(2012)]{Suzuki:2011hu} Suzuki, N., Rubin, D., Lidman, C., et al.\ 2012, \apj, 746, 85 

\bibitem[Zehavi et al.(2011)]{SDSS} Zehavi, I., Zheng, Z., Weinberg, D.~H., et al.\ 2011, \apj, 736, 59 

\bibitem[Heymans et al.(2013)]{Heymans:2013fya} Heymans, C., Grocutt, E., Heavens, A., et al.\ 2013, \mnras, 432, 2433 

\bibitem[Hildebrandt et al.(2017)]{Hildebrandt:2016iqg} Hildebrandt, H., Viola, M., Heymans, C., et al.\ 2017, \mnras, 465, 1454 

\bibitem[Burgess(2017)]{Burgess:2017ytm} Burgess, C.~P.\ 2017, arXiv:1711.10592 

\bibitem[Addison et al.(2016)]{Addison:2015wyg} Addison, G.~E., Huang, Y., Watts, D.~J., et al.\ 2016, \apj, 818, 132 

\bibitem[Riess et al.(2016)]{Riess:2016jrr} Riess, A.~G., Macri, L.~M., Hoffmann, S.~L., et al.\ 2016, \apj, 826, 56 

\bibitem[Font-Ribera et al.(2014)]{Font-Ribera:2013wce} Font-Ribera, A., Kirkby, D., Busca, N., et al.\ 2014, \jcap, 5, 027 

\bibitem[Delubac et al.(2015)]{Delubac:2014aqe} Delubac, T., Bautista, J.~E., Busca, N.~G., et al.\ 2015, \aap, 574, A59 

\bibitem[Scott(2018)]{Scott:2018adl} Scott, D.\ 2018, arXiv:1804.01318 

\bibitem[Zhao et al.(2017)]{Zhao:2017cud} Zhao, G.-B., Raveri, M., Pogosian, L., et al.\ 2017, Nature Astronomy, 1, 627

\bibitem[Sol{\`a} et al.(2017)]{runvac} Sol{\`a}, J., G{\'o}mez-Valent, A., \& de Cruz P{\'e}rez, J.\ 2017, \apj, 836, 43  

\bibitem[Di Valentino et al.(2017)]{ddetension} Di Valentino, E., Melchiorri, A., Linder, E.~V., \& Silk, J.\ 2017, \prd, 96, 023523

\bibitem[Sol{\`a} Peracaula et al.(2018)]{Sola:2017jbl} Sol{\`a} Peracaula, J., de Cruz P{\'e}rez, J., \& G{\'o}mez-Valent, A.\ 2018, \mnras, 478, 4357 

\bibitem[Ooba et al.(2018)]{Ooba:2018dzf} Ooba, J., Ratra, B., \& Sugiyama, N.\ 2018, arXiv:1802.05571 

\bibitem[Capozziello et al.(2018)]{Capozziello:2018jya} Capozziello, S., Ruchika, \& Sen, A.~A 2018, arXiv:1806.03943 

\bibitem[Das et al.(2006)]{Das:2005yj} Das, S., Corasaniti, P.~S., \& Khoury, J.\ 2006, \prd, 73, 083509 

\bibitem[Wang et al.(2015)]{Wang:2015wga} Wang, Y., Zhao, G.-B., Wands, D., Pogosian, L., \& Crittenden, R.~G.\ 2015, \prd, 92, 103005

\bibitem[Yang et al.(2018)]{Yang:2018euj} Yang, W., Pan, S., Di Valentino, E., et al.\ 2018, \jcap, 9, 019 

\bibitem[Di Valentino et al.(2017)]{idetension} Di Valentino, E., Melchiorri, A., \& Mena, O.\ 2017, \prd, 96, 043503

\bibitem[Ratra et al.(1988)]{Ratra:1987rm} Ratra B., \& Peebles P. J. E., \ 1988, \prd, 37, 3406

\bibitem[Wetterich(1995)]{Wetterich:1994bg} Wetterich, C.\ 1995, \aap, 301, 321 

\bibitem[Caldwell et al.(1998)]{Caldwell:1997ii} Caldwell, R.~R., Dave, R., \& Steinhardt, P.~J.\ 1998, Physical Review Letters, 80, 1582 

\bibitem[Steinhardt et al.(1999)]{Steinhardt:1999nw} Steinhardt, P.~J., Wang, L., \& Zlatev, I.\ 1999, \prd, 59, 123504 

\bibitem[Wetterich(2002)]{Wetterich:2001jw} Wetterich, C.\ 2002, \ssr, 100, 195 

\bibitem[Peebles \& Ratra(2003)]{PR02} Peebles, P.~J., \& Ratra, B.\ 2003, Reviews of Modern Physics, 75, 559 

\bibitem[Caldwell(2002)]{Caldwell:1999ew} Caldwell, R.~R.\ 2002, Physics Letters B, 545, 23 

\bibitem[Carroll et al.(2003)]{Carroll:2003st} Carroll, S.~M., Hoffman, M., \& Trodden, M.\ 2003, \prd, 68, 023509 

\bibitem[Vikman(2005)]{Vikman:2004dc} Vikman, A.\ 2005, \prd, 71, 023515 

\bibitem[Cline et al.(2004)]{Cline:2003gs} Cline, J.~M., Jeon, S., \& Moore, G.~D.\ 2004, \prd, 70, 043543 

\bibitem[Carroll et al.(2005)]{Carroll:2004hc} Carroll, S.~M., de Felice, A., \& Trodden, M.\ 2005, \prd, 71, 023525 

\bibitem[Wang \& Tegmark(2004)]{WT04} Wang, Y., \& Tegmark, M.\ 2004, Physical Review Letters, 92, 241302 

\bibitem[Sahni \& Starobinsky(2006)]{Sahni:2006pa} Sahni, V., \& Starobinsky, A.\ 2006, International Journal of Modern Physics D, 15, 2105 

\bibitem[Sahni et al.(2014)]{Sahni:2014ooa} Sahni, V., Shafieloo, A., \& Starobinsky, A.~A.\ 2014, \apjl, 793, L40 

\bibitem[Poulin et al.(2018)]{Poulin:2018zxs} Poulin, V., Boddy, K.~K., Bird, S., \& Kamionkowski, M.\ 2018, \prd, 97, 123504 

\bibitem[Horndeski (1974)]{Horndeski:1974wa} Horndeski, G.W., Int J Theor Phys (1974) 10: 363

\bibitem[Deffayet et al.(2009a)]{Deffayet:2009wt} Deffayet, C., Esposito-Far{\`e}se, G., \& Vikman, A.\ 2009, \prd, 79, 084003 

\bibitem[Deffayet et al.(2009b)]{Deffayet:2009mn} Deffayet, C., Deser, S., \& Esposito-Far{\`e}se, G.\ 2009, \prd, 80, 064015 

\bibitem[Deffayet et al.(2011)]{Deffayet:2011gz} Deffayet, C., Gao, X., Steer, D.~A., \& Zahariade, G.\ 2011, \prd, 84, 064039 

\bibitem[Gleyzes et al.(2015)]{Gleyzes:2014dya} Gleyzes, J., Langlois, D., Piazza, F., \& Vernizzi, F.\ 2015, Physical Review Letters, 114, 211101 

\bibitem[Zhao et al.(2009)]{Zhao:2008bn} Zhao, G.-B., Pogosian, L., Silvestri, A., \& Zylberberg, J.\ 2009, \prd, 79, 083513 

\bibitem[Hojjati et al.(2011)]{Hojjati:2011ix} Hojjati, A., Pogosian, L., \& Zhao, G.-B.\ 2011, \jcap, 8, 005 

\bibitem[Hu et al.(2014)]{Hu:2013twa} Hu, B., Raveri, M., Frusciante, N., \& Silvestri, A.\ 2014, \prd, 89, 103530 

\bibitem[Raveri et al.(2014)]{Raveri:2014cka} Raveri, M., Hu, B., Frusciante, N., \& Silvestri, A.\ 2014, \prd, 90, 043513

\bibitem[Zumalac{\'a}rregui et al.(2017)]{Zumalacarregui:2016pph} Zumalac{\'a}rregui, M., Bellini, E., Sawicki, I., Lesgourgues, J., \& Ferreira, P.~G.\ 2017, \jcap, 8, 019 

\bibitem[Damour et al.(1990)]{Damour:1990tw} Damour T., Gibbons G. W., \& Gundlach C., \ 1990, \prl, 64, 123

\bibitem[Sahni \& Shtanov(2003)]{Sahni:2002dx} Sahni, V., \& Shtanov, Y.\ 2003, \jcap, 11, 014 

\bibitem[Torres(2002)]{Torres:2002pe} Torres, D.~F.\ 2002, \prd, 66, 043522 

\bibitem[Chung et al.(2003)]{Chung:2002xj} Chung, D.~J.~H., Everett, L.~L., \& Riotto, A.\ 2003, Physics Letters B, 556, 61 

\bibitem[Faraoni(2003)]{Faraoni:2003jh} Faraoni, V.\ 2003, \prd, 68, 063508 

\bibitem[Ahmed et al.(2004)]{Ahmed:2002mj} Ahmed, M., Dodelson, S., Greene, P.~B., \& Sorkin, R.\ 2004, \prd, 69, 103523 

\bibitem[Ahmed \& Sorkin(2013)]{Ahmed:2012ci} Ahmed, M., \& Sorkin, R.~D.\ 2013, \prd, 87, 063515 

\bibitem[Zwane et al.(2017)]{Zwane:2017xbg} Zwane, N., Afshordi, N., \& Sorkin, R.~D.\ 2017, arXiv:1703.06265 

\bibitem[Crittenden et al.(2012)]{Crittenden:2011Dec} Crittenden, R.~G., Zhao, G.-B., Pogosian, L., Samushia, L., \& Zhang, X.\ 2012, \jcap, 2, 048 

\bibitem[Crittenden et al.(2009)]{Crittenden:2009Dec} Crittenden, R.~G., Pogosian, L., \& Zhao, G.-B.\ 2009, \jcap, 12, 025 

\bibitem[Wang \& Dai(2016)]{Wang:2015tua} Wang, Y., \& Dai, M.\ 2016, \prd, 94, 083521 

\bibitem[Planck Collaboration et al.(2018)]{PLC18} Planck Collaboration, Aghanim, N., Akrami, Y., et al.\ 2018, arXiv:1807.06209 

\bibitem[Betoule et al.(2014)]{Betoule:2014frx} Betoule, M., Kessler, R., Guy, J., et al.\ 2014, \aap, 568, A22 

\bibitem[Beutler et al.(2011)]{6df} Beutler, F., Blake, C., Colless, M., et al.\ 2011, \mnras, 416, 3017 

\bibitem[Ross et al.(2015)]{MGS} Ross, A.~J., Samushia, L., Howlett, C., et al.\ 2015, \mnras, 449, 835 

\bibitem[Wang et al.(2017)]{BAOWang} Wang, Y., Zhao, G.-B., Chuang, C.-H., et al.\ 2017, \mnras, 469, 3762 

\bibitem[Ata et al.(2018)]{Ata:2017dya} Ata, M., Baumgarten, F., Bautista, J., et al.\ 2018, \mnras, 473, 4773 

\bibitem[Moresco et al.(2016)]{Moresco:2016mzx} Moresco, M., Pozzetti, L., Cimatti, A., et al.\ 2016, \jcap, 5, 014 

\bibitem[Handley et al.(2015a)]{polychordI} Handley, W.~J., Hobson, M.~P., \& Lasenby, A.~N.\ 2015, \mnras, 450, L61

\bibitem[Handley et al.(2015b)]{polychordII} Handley, W.~J., Hobson, M.~P., \& Lasenby, A.~N.\ 2015, \mnras, 453, 4384 

\bibitem[Lewis \& Bridle(2002)]{cosmomc} Lewis, A., \& Bridle, S.\ 2002, \prd, 66, 103511  

\bibitem[Lewis et al.(2000)]{CAMB} Lewis, A., Challinor, A., \& Lasenby, A.\ 2000, \apj, 538, 473 

\bibitem[Lemos et al.(2018)]{LLEG18} Lemos, P., Lee, E., Efstathiou, G., \& Gratton, S.\ 2018, arXiv:1806.06781 

\bibitem[Abbott et al.(2017)]{Abbott:2017xzu} Abbott, B.~P., Abbott, R., Abbott, T.~D., et al.\ 2017, \nat, 551, 85 

\bibitem[Guidorzi et al.(2017)]{Guidorzi:2017ogy} Guidorzi, C., Margutti, R., Brout, D., et al.\ 2017, \apjl, 851, L36 

\bibitem[Hotokezaka et al.(2018)]{Hotokezaka:2018dfi} Hotokezaka, K., Nakar, E., Gottlieb, O., et al.\ 2018, arXiv:1806.10596 

\bibitem[Bergmann(1968)]{Bergmann1968} Bergmann P. G., \ 1968, International Journal of Theoretical Physics, 1, 25

\bibitem[Nordtvedt(1970)]{Nordtvedt1970} Nordtvedt, K., Jr.\ 1970, \apj, 161, 1059 

\bibitem[Wagoner(1970)]{Wagoner1970} Wagoner R. V.,\ 1970 \prd, 1, 3209  

\bibitem[Zhao et al.(2017)]{BAOZhao} Zhao, G.-B., Wang, Y., Saito, S., et al.\ 2017, \mnras, 466, 762 

\end{thebibliography}
\end{document}